\documentclass[conference]{IEEEtran}
\IEEEoverridecommandlockouts
\usepackage{cite,color}
\usepackage{amsmath,amssymb,amsfonts}

\usepackage{algpseudocode}

\usepackage[ruled,vlined]{algorithm2e}
\usepackage{graphicx}
\usepackage{textcomp}
\usepackage{bm}
\usepackage{xcolor}
\def\BibTeX{{\rm B\kern-.05em{\sc i\kern-.025em b}\kern-.08em
    T\kern-.1667em\lower.7ex\hbox{E}\kern-.125emX}}
\begin{document}

\title{Diff-GO$^\text{n}$: Enhancing Diffusion Models for Goal-Oriented Communications
\vspace*{-4mm}}

% \title{  Diffusion Model Enhancements for Goal-Oriented Communication \\

% % \thanks{Identify applicable funding agency here. If none, delete this.}
% }

\author{\IEEEauthorblockN{ Suchinthaka Wanninayaka$^1$, Achintha Wijesinghe$^1$, Weiwei Wang$^1$, Yu-Chieh Chao$^1$, \\
Songyang Zhang$^2$,~\textit{Member, IEEE} and
Zhi Ding$^1$, \textit{Fellow,~IEEE}}
\IEEEauthorblockA{$^1$University of California at Davis
$\qquad$
$^2$University of Louisiana at Lafayette}}

\maketitle

\begin{abstract}
The rapid expansion of edge devices and Internet-of-Things (IoT) continues to heighten the demand for data transport 
under limited spectrum resources. 
The goal-oriented communications (GO-COM),
unlike traditional communication systems designed for bit-level accuracy,  prioritizes more critical information for specific
application goals at the receiver.  
To improve the efficiency of
generative learning models for GO-COM, this work introduces a novel
noise-restricted diffusion-based GO-COM (Diff-GO$^\text{n}$) framework 
for reducing bandwidth overhead while preserving the
media quality at the receiver. 
Specifically, we propose an innovative Noise-Restricted Forward Diffusion (NR-FD) framework to
accelerate 
model training and reduce the computation burden for diffusion-based GO-COMs by 
leveraging a pre-sampled pseudo-random
noise bank (NB). Moreover, we design an early stopping criterion for improving computational efficiency and convergence speed, allowing high-quality generation in fewer training steps. Our experimental results demonstrate superior perceptual quality of data transmission 
at a reduced bandwidth usage and lower computation, making Diff-GO$^\text{n}$ 
well-suited for real-time communications and downstream applications.
\end{abstract}


\section{Introduction}
Future-generation wireless communications, such as emerging sixth-generation (6G) standard cellular networks, 
are expected to support a broad range of edge devices and services, demanding the capability to ensure the transmission of large volumes of data traffic with limited bandwidth and constrained computation resources \cite{9349624}. Traditional data-oriented communications (DO-COM), inherited from Shannon's communication model, focus on packet transmission and bit-wise recovery and reach a performance limit bounded by the Shannon capacity \cite{10570351}. However, in some scenarios, the exact bit-wise message recovery is unnecessary to accomplish the task at the receiver end. For example, the environmental information, such as the depth, directions, and size of nearby objects, is more important than the visual details, such as color and window size, in autonomous driving \cite{Diff-go}. Thus, with the development of artificial intelligence (AI), Goal-Oriented Communication (GO-COM) emerges as an important concept in the next-generation wireless communications.
Unlike traditional GO-COMs, GO-COM systems prioritize transmitting information that is most critical to the end task. By targeting the delivery of the key semantic messages for downstream tasks, GO-COMs facilitate more efficient data transmission, optimizing bandwidth usage while maintaining overall system performance, particularly in resource-constrained applications.

The concept of GO-COM stems from deep learning for communications, where the message
transmission system is formulated as a deep neural network (DNN) based autoencoders \cite{o2017introduction}. The aforementioned DNN-based models are inspired by several advanced generative frameworks, including SPADE \cite{SPADE}, OASIS \cite{OASIS} and CC-FPSE \cite{CC-FPSE}, succeeding in the efficient image generation.
However, these autoencoder-based communication systems lack interpretation of
insightful meaning of raw data, thereby leading to redundant communication overhead and less
compression rate. 
Recent developments in generative AI are
fueling exciting technologies, including the diffusion model \cite{croitoru2023diffusion}, which has emerged as a more effective network backbone for GO-COMs. Typical diffusion-based GO-COMs include GESCO\cite{GESCO}, Diff-GO \cite{Diff-go}, and Diff-GO+ \cite{Diff-go+}, which utilize Denoising Diffusion Probabilistic Models (DDPM)\cite{DDPM} to embed high-fidelity semantic representations and regenerate data.
Specifically, GESCO\cite{GESCO} generates images without transmitting \footnote{We term the forward diffusion output ``noised latent", following \cite{improved}. }{noised latent} from forward diffusion, whereas Diff-GO\cite{Diff-go} and Diff-GO+\cite{Diff-go+} improve the reconstruction quality by incorporating noised latent. Despite some successes, these diffusion-based frameworks suffer from extensive training time and high computational demands. Moreover, the inclusion of noised latent further increases communications overhead, limiting their efficiency in real-time applications. Thus, leveraging the diffusion model's power in GO-COMs while reducing communication and computation costs remains a critical research focus.

\begin{figure*}[t]
\centering
\includegraphics[height=0.235\textwidth, width=0.93\textwidth]{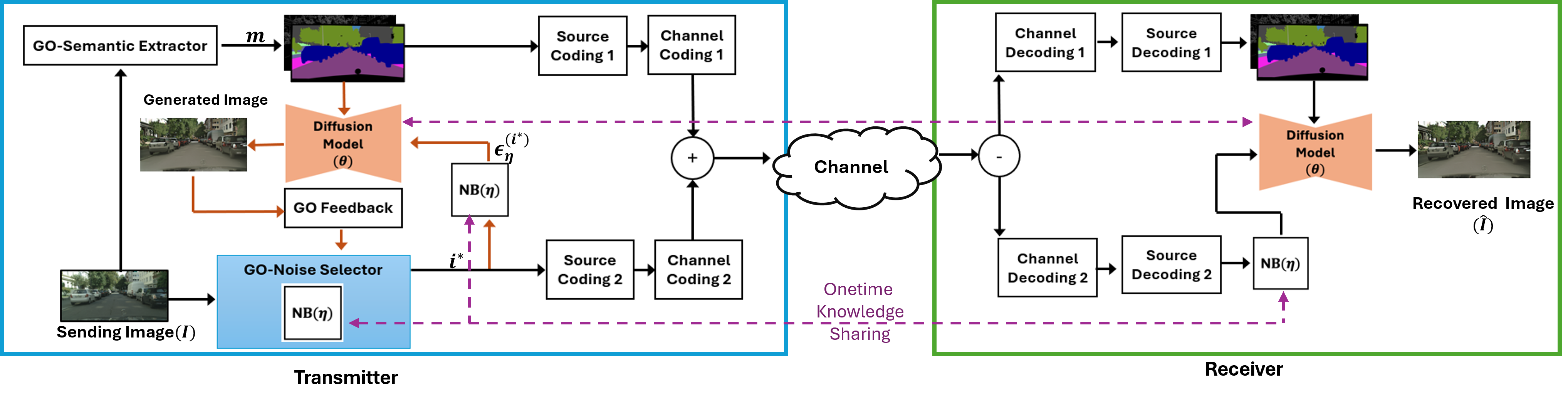}
\vspace{-4mm}
\caption{Overall architecture of {Diff-GO}$^\text{n}$: a) Transmitter (TX) Sides - Extract semantic conditions and represent the noised latent from forward diffusion by noise banks; b) Receiver (RX) Side - Reconstruct the noised latent and implement the diffusion model for image regeneration conditioned on semantic features.} 
\label{fig: System_architecture}
\vspace{-3mm}
\end{figure*}

To this end, we propose a novel framework of \textbf{n}oise-restricted \textbf{Diff}usion-based \textbf{GO}-COM, namely \textit{Diff-GO}$^{n}$, which aims to facilitate diffusion model with efficient training and latent noise transmission. Specifically, we introduce
a Noise-Restricted Forward Diffusion (NR-FD) design 
which incorporates a pre-sampled collection of noise vectors—referred to as noise banks (NB)—into the traditional forward diffusion process. Instead of transmitting the
high entropy latent noise
representation to the receiver,
the NR-FD 
drastically reduces the communication overhead and improves the reconstruction accuracy at the receiver by only sending the indices of the spanning noise vectors within NB.
Furthermore, we propose an early stopping criterion in training, guided by the Learned Perceptual Image Patch Similarity (LPIPS) score \cite{zhang2018unreasonable}, to determine the optimal stopping point for diffusion training. Our contributions are summarized as follows:
\begin{itemize}
  \item We propose Diff-GO$^\text{n}$: a novel  framework for GO-COMs based on an innovative design of Noise-Restricted Forward Diffusion (NR-FD) with a noise bank, 
which significantly reduces communication bandwidth while still achieving a superior 
reception quality.
\item To lower the training computational cost of diffusion models, 
we provide an early stopping criterion based on the LPIPS score, 
which halts the diffusion process upon reaching the desired reconstruction quality.

\item We perform extensive experiments on the proposed Diff-GO$^\text{n}$
in comparison with existing state-of-the-art (SOTA) semantic communication frameworks.
We test the cityscape dataset for image transmission 
to establish benchmark results. The demonstrated superior performance validates 
the power of NR-FD and the efficiency of our Diff-GO$^\text{n}$, 
particularly under limited communication and computational resources.
\end{itemize}

% \begin{figure*}[htbp]
% \centering
% \begin{minipage}[b]{0.5\linewidth}
% \centering
% \includegraphics[width=1.2\columnwidth]{Figures/training.png}
% \caption{}
% \label{fig:training_framework}
% \end{minipage}\vspace*{-5mm}
% \end{figure*}

\section{Overall Architecture} \label{sec:Overall Architecture}
In this section, we first introduce the overall architecture of the proposed Diff-GO$^\text{n}$ communication system
captured in Fig. \ref{fig: System_architecture}.  We then present the
detailed functional blocks in Section \ref{sec:Methodology}. 
Conceptually, our Diff-GO$^\text{n}$ consists of two phases: the training phase and the inference (showtime) phase. 
Diff-Go$^\text{n}$ features a novel NB and a GO-Training Controller with a newly designed NR-FD structure. For convenience, we will use image transmission as a primary example, which is useful for many practical applications, such as remote sensing/control,
autonomous driving, and traffic monitoring.

\subsection{Training Phase of Diff-GO$^\text{n}$}
\begin{figure}[t]
\vspace*{-4mm}
\centering
\includegraphics[width=0.38\textwidth]{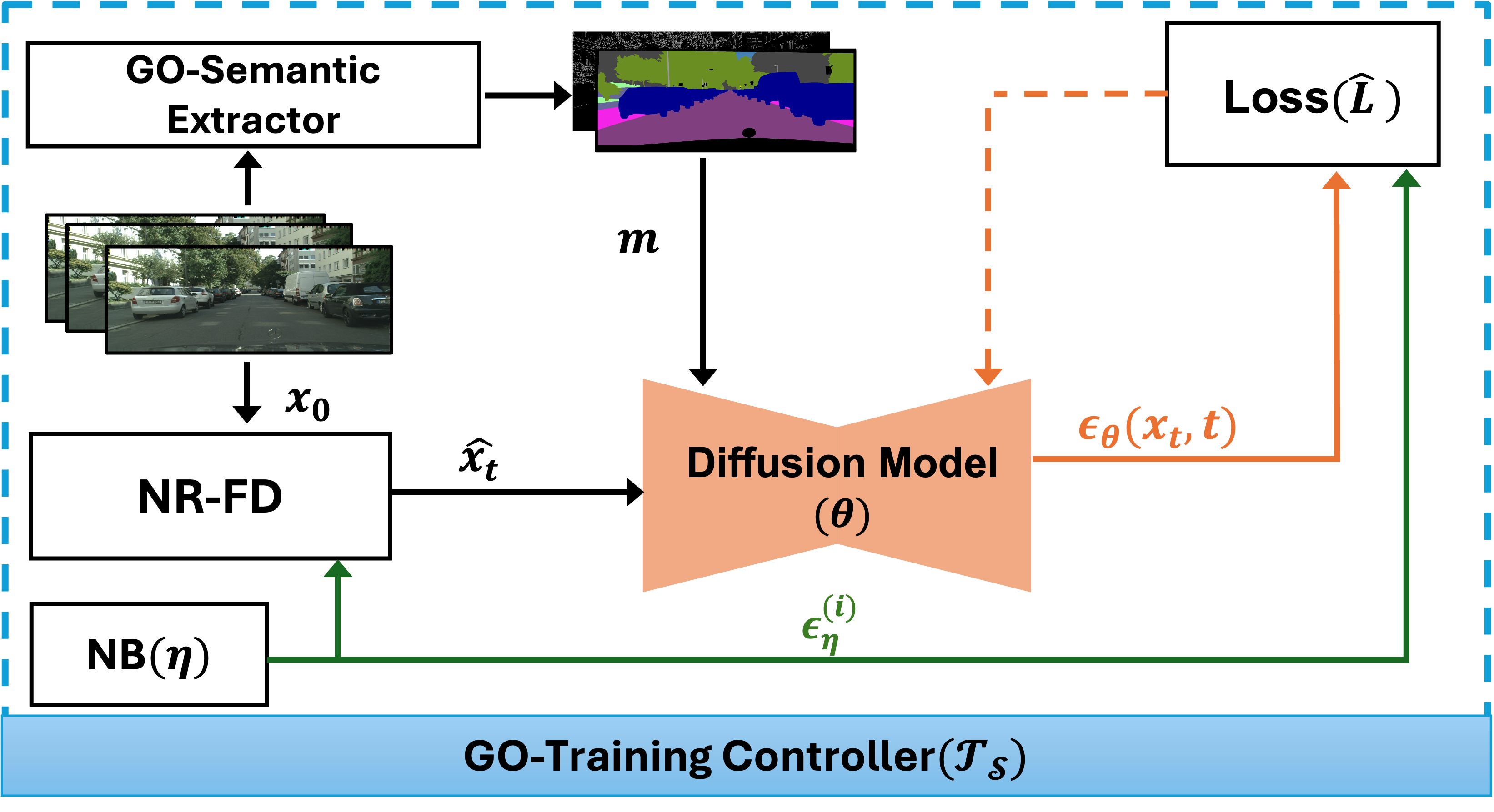}
\vspace*{-4mm}
\caption{Training process of Diff-GO$^\text{n}$.}
\label{fig:training_framework}
\vspace*{-6mm}
\end{figure}

The training phase of Diff-GO$^\text{n}$ aims to efficiently train a diffusion model conditioned on the key semantic goal-oriented (GO) information using a structured process shown in Fig. \ref{fig:training_framework}. In particular, this work utilizes 
the well-known DDPM \cite{DDPM}, which progressively adds pseudorandom
noise to the original data through a forward diffusion process until the noisy data
approaches an isotropic Gaussian distribution. 
The model then reverses to recover the original data by reversing this process via a learned denoising function. To enable reverse image synthesis (i.e., recovery) in Diff-GO$^\text{n}$, 
the noised latent representation from forward diffusion, together with semantic conditions 
are fed to the train model.
In the training process, the source image $\mathbf{x}_0$ first passes through the GO-Semantic Extractor, which generates semantic conditions $\mathbf{m}$ to characterize the critical information about the 
source data for downstream tasks. Since we consider the example of
image transmission for remote-controlled vehicles 
where object size and distance are more important, 
typical examples of semantic conditions include the segmentation map, depth, and edge map. 

At each diffusion step of the forward diffusion process, instead of randomly sampling noise from the standard Gaussian distribution, $\mathcal{N}( \mathbf{0}, \mathbf{I})$, we select a noise vector $\boldsymbol{\epsilon}_ \eta^{(i)}$ from the Noise Bank (NB). The NB is constructed from a pre-sampled collection of noise vectors, denoted by $\boldsymbol{\eta}$. 
In each NR-FD training step, the selected noise vector is applied to generate a noised latent representation $\hat{\mathbf{x}}_t$, which is then passed to the diffusion model. 
Subsequently, the diffusion model updates its parameters($\theta$) by predicting the noise $\boldsymbol{\epsilon}_ \theta\left(\mathbf{x}_t, t\right)$ added during the forward diffusion step, t. 
The model compares its prediction with the actual noise through a loss function $\hat{L}$
to refine the model's performance.
The details of the training and loss function will be discussed in Section \ref{sec:noise}.

To ensure that the model converges efficiently, the GO-Training Controller monitors the quality of the image reconstructions throughout the training process. The controller calculates the LPIPS score at regular intervals to assess the perceptual similarity between the generated and original images. If the score meets a pre-defined threshold $\mathcal{T}_{S}$, the controller triggers early stopping, effectively ending the training process and reducing computational overhead.

% \subsection{Noise-Restricted and Communication}
\subsection{Communication Phase of Diff-GO$^\text{n}$}
Once the training of the diffusion model is over, the trained model is deployed at both the transmitter and receiver end, we enter the communication (or “showtime”) phase. The showtime phase
begins with generating semantic conditions, $\mathbf{m}$ in the GO-Semantic Extractor. 
To ensure successful reconstruction or synthesis of the source
image through reverse diffusion at the receiver end, the noised latent representation
from forward diffusion, together with the semantic conditions $\mathbf{m}$ 
are transmitted by the transmitter. 
However, the noised latent representation has high entropy, making it challenging to compress and transmit efficiently. However, the direct encoding and transmission of typically
high-entropy noised latent representation would
lead to remarkably high data payload, consuming massive network bandwidth.
To this end, we design a GO-Noise Selector (GONS) to 
select a noise vector $\boldsymbol{\epsilon}_ \eta{\left(i^*\right)}$ from the NB
according to the basis of Gaussian radius matching, 
with details to be presented in Section \ref{sec:opt}. 
Once the noise vector is selected, the transmitter
only needs to encode the index $i^*$ 
instead of the high entropy latent representation, 
thereby significantly reducing the cost of spectrum and power resources. 

To efficiently communicate the indexed noised latent presentation and 
the extracted semantic feature conditions, we devise a customized dual-level coding strategy. 
Specifically, we utilize two distinct coding rates
for the semantic conditions and the noise index, respectively, as shown in Fig. \ref{fig: System_architecture}. We apply a more robust encoding for stronger
error protection of semantic conditions, and we provide a
weaker error correction for the NB index transmission.
This approach is more efficient since generative learning models
tend to be less sensitive to modest corruption of the noise
representation. 

It is important to note that the NB is pre-configured and shared between
the transmitter and the receiver (one-time knowledge sharing)  after training and prior to real-time communication in showtime phase. 
After decoding the transmitted semantic conditions and noise index at the receiver, these elements are jointly used to initiate image generation. The semantic conditions $\mathbf{m}$ serve as the conditioning input, while the noise $\boldsymbol{\epsilon}_ \eta{\left(i^*\right)}$ provides the starting point for the diffusion model, which progressively denoises $\boldsymbol{\epsilon}_ \eta{\left(i^*\right)}$ over multiple steps to reconstruct the original image.

\section{Detailed Network Structure}\label{sec:Methodology}
\vspace*{-0.5mm}

In this section, we present the mathematical formulation and training process of our Diff-GO$^\text{n}$ framework, beginning with a brief overview of DDPM.

\vspace*{-0.5mm}
\subsection{Denoising Diffusion Probabilistic Models}
Our basic Diff-GO$^\text{n}$ architecture centers around the DDPM model, consisting of the well-known forward and reverse bidirectional diffusion process.

The forward diffusion process successively adds Gaussian noises to the $\mathbf{x}_0 \sim q\left(\mathbf{x}_0\right)$ spanning $T$ discrete time steps.
Denoted by $q\left(\mathbf{x}_{1:T} \mid \mathbf{x}_0\right)$, 
the forward diffusion process manifests through the following equations:
\begin{align}
   q(\mathbf{x}_{1:T}|\mathbf{x}_0) &= \prod_{t=1}^{T} q(\mathbf{x}_t|\mathbf{x}_{t-1}),\\
   q(\mathbf{x}_t|\mathbf{x}_{t-1}) & = \mathcal{N}(\mathbf{x}_t;\sqrt{1-\beta_t}\mathbf{x}_{t-1},\beta_t  \mathbf{I}),
    \label{eqn:fwd}
\end{align}
where $\beta_t$ and $\mathbf{x}_t$ represent the variance of the noise and the noisy sample at the time step $t$, respectively. A notable property of the forward diffusion process is that the noisy sample $\mathbf{x}_t$ at any time step $t$ can be calculated directly from the $\mathbf{x}_0$
and can be expressed as \vspace*{-2mm}
\begin{align}
    q\left(\mathbf{x}_t \mid \mathbf{x}_0\right)&=\mathcal{N}\left(\mathbf{x}_t ; \sqrt{\bar{\alpha}}_t \mathbf{x}_0,\left(1-\bar{\alpha}_t\right) \mathbf{I}\right), \\
    \mathbf{x}_t&=\sqrt{\bar{\alpha}_t} \mathbf{x}_0+\sqrt{1-\bar{\alpha}_t} \boldsymbol{\epsilon},\label{eq1}
\end{align}
where $\alpha_t:=1-\beta_t$,  $\bar{\alpha}_t:=\prod_{l=0}^t \alpha_l$ and $ \boldsymbol{\epsilon} \sim \mathcal{N}( \mathbf{0}, \mathbf{I})$.

The reverse diffusion process aims to recover the $\mathbf{x}_0$ by progressive denoising (removing the noise introduced during the forward process). Suppose that $\mathbf{x}_T \sim \mathcal{N}( \mathbf{0}, \mathbf{I})$. The reverse process is modeled by the following joint probability density
notations:\vspace*{-2mm}
\begin{align}
p_\theta\left(\mathbf{x}_{0: T}\right)&=p\left(\mathbf{x}_T\right) \prod_{t=1}^T p_\theta\left(\mathbf{x}_{t-1} \mid \mathbf{x}_t\right),\\
p_\theta\left(\mathbf{x}_{t-1} \mid \mathbf{x}_t\right)&=\mathcal{N}\left(\mathbf{x}_{t-1} ; \boldsymbol{\mu}_ \theta\left(\mathbf{x}_t, t\right), \boldsymbol{\Sigma}_ \theta\left(\mathbf{x}_t, t\right)\right).
\end{align}

Here, $\boldsymbol{\mu}_ \theta\left(\mathbf{x}_t, t\right)$ and $\boldsymbol{\Sigma}_ \theta\left(\mathbf{x}_t, t\right)$ represent the predicted mean and variance for each reverse transition. The reverse process progressively denoises $\mathbf{x}_T$ until it probabilistically approaches
the $\mathbf{x}_0$. The noise addition $\epsilon$ during the forward diffusion can be learned via a 
deep neural network $\boldsymbol{\epsilon}_ \theta\left(\mathbf{x}_t, t\right)$ with parameters $\theta$ by minimizing the following loss function:
\begin{equation}
L=\mathbb{E}_{t, \mathbf{x}_0, \boldsymbol{\epsilon}}\left[\left\|\boldsymbol{\epsilon}-\boldsymbol{\epsilon}_ \theta\left(\mathbf{x}_t, t\right)\right\|^2\right]
\end{equation}
This objective allows the model to accurately predict the noise at each time step, thereby leading to an effective reconstruction of the $\mathbf{x}_0$ from $\mathbf{x}_T$.

\vspace*{-1mm}

\subsection{Noise-Restricted Diffusion Model} \vspace*{-1mm}
\label{sec:noise}
As introduced in the previous section, instead of pseudo-randomly generating noised latent representation without control, we customize the forward diffusion process by constructing a pre-sampled noise bank for diffusion models. 
We propose to construct a NB from a set of 
Gaussian noise sample vectors
$\boldsymbol{\eta}=\left\{\boldsymbol{\epsilon}_ \eta {(1)}, \boldsymbol{\epsilon}_ \eta {(2)}, \ldots,\boldsymbol{\epsilon}_ \eta{(N)}\right\}$, each $\boldsymbol{\epsilon}_ \eta {(i)}$ of which is randomly drawn from
$\mathcal{N}( \mathbf{0}, \mathbf{I})$. The size of the NB is $N$.
The use of a pre-sampled NB introduces a degree
of control to the pseudo-random noise generation in the forward diffusion process. Such controlled
sampling has the potential to improve the 
diffusion model performance in GO-COM applications.

In our customized forward diffusion process of Noise-Restricted Forward Diffusion (NR-FD), we select a noise vector $\boldsymbol{\epsilon}_ \eta {(i)}$ from the NB, thereby
modifying Eq. \eqref{eq1} into
\begin{equation}
\hat{\mathbf{x}}_t=\sqrt{\bar{\alpha_t}} \mathbf{x}_0+\sqrt{1-\bar{\alpha}_t} \boldsymbol{\epsilon}_ \eta {(i)},
\end{equation}
where $\eta$ indicates the specific NB collection and $i$ identifies the instance of noise vector from NB selected at each step. This introduction of restricted noise ensures that the model can expect more predictable noise patterns and strengthens the efficiency of the reverse diffusion process.

The reverse diffusion process is adapted to predict the specific noise vector $\boldsymbol{\epsilon}_ \eta {(i)}$. Unlike traditional DDPM, the neural network $\boldsymbol{\epsilon}_ \theta\left(\mathbf{x}_t, t\right)$ is trained to predict the specific noise vector $\boldsymbol{\epsilon}_ \theta\left(\mathbf{x}_t, t\right)$ from the NB rather than an independently sampled Gaussian noise, which can be characterized as
\begin{equation}
\hat{L}=\mathbb{E}_{t, \mathbf{x}_0, \boldsymbol{\epsilon}_ \eta {(i)}}\left[\left\|\boldsymbol{\epsilon}_ \eta {(i)}-\boldsymbol{\epsilon}_ \theta\left(\mathbf{x}_t, t\right)\right\|^2\right]
\end{equation}

This formulation further supports accurate and reliable reconstructions by focusing on the limited noise patterns from the NB.

\vspace{0.1mm}
\setlength{\textfloatsep}{2pt}% Remove \textfloatsep
\begin{algorithm}[t]
    \caption{Training Phase of Diff-GO$^\text{n}$}
    \label{alg:training}
    \DontPrintSemicolon
    \LinesNumbered
    \SetNlSty{}{}{:}
    \KwIn{Training data set: $\mathcal{D}$}
    \KwIn{Number of noise vectors in the Noise Bank: $N$}
    \KwIn{Target similarity score: $\mathcal{T}_{S}$}
    \KwIn{Validation set size: $S$}
    \KwIn{Check interval: $\kappa$}
    \KwOut{Trained diffusion model: $\mathcal{\theta}$}
    \KwOut{Noise Bank: $\bm{\eta}$}
    Initialize Noise Bank $\bm{\eta}$: $\bm{\eta} \xleftarrow{} \{\boldsymbol{\epsilon}_ {\eta}{(1)}, \ldots, \boldsymbol{\epsilon}_ {\eta}{(N)} \}$\\
    Initialize iteration counter: $\lambda \xleftarrow{} 0$\\
    \While{$\mathcal{\theta}$ is not converged}{
        \For{each $\mathbf{d} \in \mathcal{D}$}{
            Sample $\boldsymbol{\epsilon}_ {\eta}{(i)}$ from $\bm{\eta}$\\
            $\hat{\mathbf{x}_t} \xleftarrow{} \sqrt{\bar{\alpha}_t} \mathbf{d} + \sqrt{1 - \bar{\alpha}_t} \boldsymbol{\epsilon}_ {\eta}{(i)}$ (NR-FD)\\
            Train model $\mathcal{\theta}$: $\mathcal{\theta} \xleftarrow{}$ Update using $L = \mathbb{E}_{t, \mathbf{d}, \boldsymbol{\epsilon}_ {\eta}{(i)}}\left[\left\|\boldsymbol{\epsilon}_ {\eta}{(i)} - \boldsymbol{\epsilon}_ \theta(\hat{\mathbf{x}_t}, t)\right\|^2 \right]$\\
            $\lambda \xleftarrow{} \lambda + 1$\\
        }
        \If{$\lambda \% \kappa == 0$}{
            Generate a validation set of images: $\mathcal{I} = \{I_1, I_2, \dots, I_S\}$\\
            Compute similarity score: $\mathcal{S}_{\kappa} \xleftarrow{} \text{Sim}(\mathcal{I})$\\
            \If{$\mathcal{S}_{\kappa} \leq \mathcal{T}_{S}$}{
                Stop diffusion process\\
                \textbf{break}
            }
        }
    }
    Share trained model $\mathcal{\theta}$ and NB $\bm{\eta}$ with receiver
\end{algorithm}

\vspace{0.1mm}
\begin{algorithm}[t]
    \caption{TX Side: Noise and Semantic Extraction}
    \label{alg:transmission}
    \DontPrintSemicolon
    \LinesNumbered
    \SetNlSty{}{}{:}
    \KwIn{Inference image: $\mathcal{I}$}
    \KwIn{Condition: $\mathbf{m}$ (extracted from the image)}
    \KwIn{Theoretical Gaussian radius: $r_{\phi}$}
    \KwIn{Noise Bank: $\bm{\eta} = \{\boldsymbol{\epsilon}_ {\eta}{(1)}, \boldsymbol{\epsilon}_ {\eta}{(2)}, \dots, \boldsymbol{\epsilon}_ {\eta}{(N)}\}$}
    \KwOut{Selected noise index from Noise Bank: $i^*$}
    \KwOut{Condition: $\mathbf{m}$}

    $\mathbf{m} \xleftarrow{} \text{extract condition from } \mathcal{I}$ \Comment{Extract semantic meaning or condition from the image} \\

    \For{each $\boldsymbol{\epsilon}_ {\eta}{(i)} \in \bm{\eta}$}{
        $\mathbf{x}_T^{(i)} \xleftarrow{} \sqrt{\bar{\alpha}_T} \mathbf{x}_0 + \sqrt{1 - \bar{\alpha}_T} \boldsymbol{\epsilon}_ {\eta}{(i)}$ \Comment{Forward diffusion at step $T$ using $\boldsymbol{\epsilon}_ {\eta}{(i)}$ from the Noise Bank}\\
        $r_{\psi}^{(i)} \xleftarrow{} \text{Gaussian radius of } \mathbf{x}_T^{(i)}$ \Comment{Calculate Gaussian radius for $\boldsymbol{\epsilon}_ {\eta} {(i)}$}\\
    }

    $i^* \xleftarrow{} \arg \min_{i} |r_{\psi}^{(i)} - r_{\phi}|$ \Comment{Select the noise vector that best matches the theoretical Gaussian radius}\\
    
    \textbf{Output:} Send the selected noise index $i^*$ and the condition $\mathbf{m}$

\end{algorithm}

\begin{algorithm}[htb]
    \caption{RX Side: Image Regeneration}
    \label{alg:Regeneration}
    \DontPrintSemicolon
    \LinesNumbered
    \SetNlSty{}{}{:}
    \KwIn{Noise index: $i^*$}
    \KwIn{Condition: $\mathbf{m}$ (received from sender)}
    \KwIn{Noise Bank: $\bm{\eta} = \{\boldsymbol{\epsilon}_ {\eta}{(1)}, \boldsymbol{\epsilon}_ {\eta}{(2)}, \dots, \boldsymbol{\epsilon}_ {\eta}{(N)}\}$}
    \KwIn{Trained model: $\mathcal{\theta}$}
    \KwOut{Regenerated image: $\hat{\mathcal{I}}$}

    Retrieve noise vector: $\boldsymbol{\epsilon}_ {\eta}{(i^*)} \xleftarrow{} \bm{\eta}[i^*]$ \Comment{Fetch the noise from the Noise Bank based on received index $i^*$}\\

    Initialize $\mathbf{x}_T \xleftarrow{} \sqrt{\bar{\alpha}_T} \mathbf{x}_0 + \sqrt{1 - \bar{\alpha}_T} \boldsymbol{\epsilon}_ {\eta}{(i^*)}$ \Comment{Initialize with forward-diffused noise using the selected noise vector}\\

    \For{$t = T$ \textbf{to} $1$}{
        $\mathbf{x}_{t-1} \xleftarrow{} \mathcal{\theta}(\mathbf{x}_t, \mathbf{m})$ \Comment{Reverse diffusion step using the model $\mathcal{\theta}$ and the condition $\mathbf{m}$}\\
        $\mathbf{x}_t \xleftarrow{} \mathbf{x}_{t-1}$
    }

    $\hat{\mathcal{I}} \xleftarrow{} \mathbf{x}_0$ \Comment{Final reconstructed image after reverse diffusion}

    \textbf{Output:} Regenerated image $\hat{\mathcal{I}}$
\end{algorithm}

\vspace*{-2.5mm}
\addtolength{\topmargin}{0.1in}
\subsection{Transmission of Noised Latent via NB} \vspace*{-1mm}\label{sec:opt}
Transmitting the latent representation $\mathbf{x}_T$ increases
data payload and consumes additional bandwidth overhead. To further improve spectrum efficiency, we only need
to encode the index $i^*$ of the selected noise vector $\boldsymbol{\epsilon}_ \eta{\left(i^*\right)}$ from the $\boldsymbol{\eta}$. Specifically, the noised latent $\mathbf{x}_T^{(i)}$ at the final forward diffusion step $T$ can be calculated by
\vspace*{-2mm}
\begin{equation}
\mathbf{x}_T^{(i)}=\sqrt{\bar{\alpha}_T} \mathbf{x}_0+\sqrt{1-\bar{\alpha}_T} \boldsymbol{\epsilon}_ \eta{(i^*)}.
\end{equation}

To select the suitable noise vector $\boldsymbol{\epsilon}_ \eta {\left(i^*\right)}$ , we compute its Gaussian radius $r_\psi^{(i)}$ compared to the theoretical Gaussian radius $r_\phi$ , following the approach in \cite{radius}, to minimize the distance \vspace*{-2mm}
\begin{equation}
i^*=\arg \min _i\left\|r_\psi^{(i)}-r_\phi\right\|,
\end{equation}
where $i^*$ is the index of the best-matching noise vector. By encoding only this noise index $i^*$ for transmission, 
we significantly lower the channel payload while still ensuring high-quality reconstruction at the receiver end.

\subsection{Early Stopping for Efficient Training}
One major shortcoming in diffusion-based GO-COM is the substantial computational complexity required to train a
satisfactory model.
For example, classic diffusion-based frameworks, such as Diff-GO and Diff-GO+, require around 250,000 steps to reach their near-optimum performance. Such a lengthy training process not only consumes a lot of computational resources but also hampers the implementation of practical communication. To mitigate this concern, we introduce 
an early stopping criterion based on the LPIPS score, 
one of the commonly used metrics for measuring perceptual similarity. This mechanism allows the model to 
terminate training once a certain quality threshold is achieved, thereby leading to faster convergence without sacrificing performance.
This early stopping process can be described as follows. During the training, noise vectors $\boldsymbol{\epsilon}_ \eta {(i)}$ from the NB are pseudo-randomly drawn to perform forward diffusion on each data sample $d$. 
At pre-defined intervals determined by the check interval $\kappa$, a validation set of images 
$\mathcal{I}=\left\{I_1, I_2, \ldots, I_S\right\}$ are applied
to generate the LPIPS score $\mathcal{S}_{\kappa}$ to 
evaluate the Diff-GO$^{n}$ quality. Once the LPIPS score 
falls below the target threshold $\mathcal{T}_{S}$, the training process is stopped earlier. This local feedback approach can dynamically adjust the training steps for different datasets, which would adaptively speed up
the training process based on the quality of generated images. 

To provide a better illustration of our Diff-GO$^\text{n}$, our training and inference steps are described in \textbf{Algorithm 1-3}.

\section{Experiments}\label{sec:Results And Discussion}
We now present our empirical results in comparison with 
several SOTA frameworks. To demonstrate the efficacy of the proposed Diff-GO$^\text{n}$, we present the results using the Cityscape dataset \cite{Cityscapes}, which contains over
5000 annotated high-resolution image frames of urban street scenes from fifty cities. We focus on the Cityscape dataset since many Diff-GO-COM frameworks also work with this dataset, making direct comparison very expedient. 
\vspace*{-2mm}

\subsection{Quality of Reconstructed Message on the Receiver}
\begin{figure}[t]
%\vspace*{-5mm}
\centerline{\includegraphics[width=0.35\textwidth]{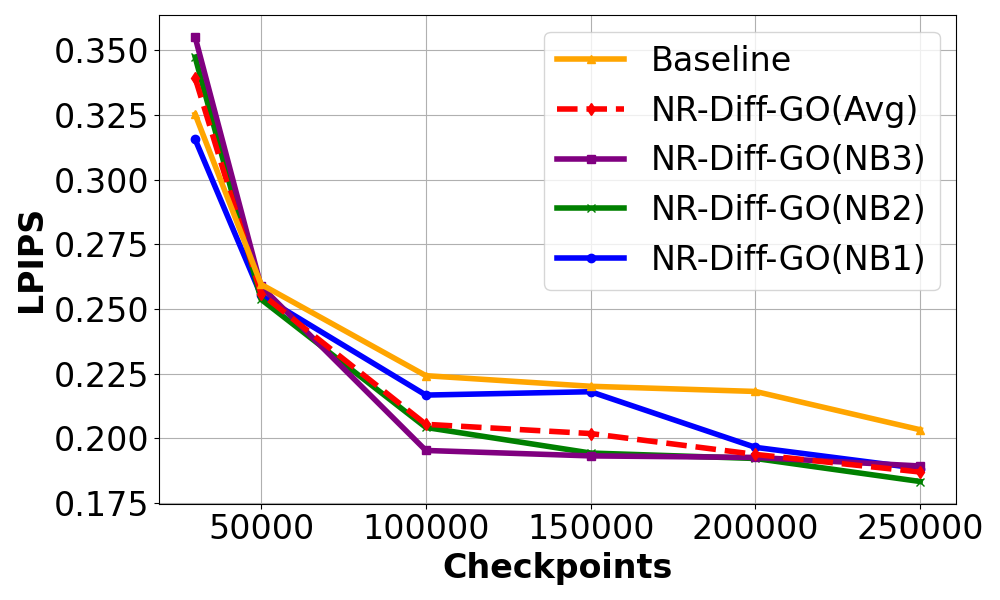}}
\vspace*{-3mm}
\caption{Comparison of Diff-GO$^\text{n}$ with DDPM (Baseline) under different NB initialization by using LPIPS scores.}
\label{fig:convergence}
\vspace*{-1mm}
\end{figure}

% \begin{figure}[t]
% %\vspace*{-4mm}
% \centerline{\includegraphics[width=0.35\textwidth]{Figures/ADE20k.png}}
% \vspace*{-3mm}
% \caption{Comparison between SOTA GO-COMs and Diff-GO$^\text{n}$ in ADE2 0K.}
% \label{fig:ADE20K}
% \vspace*{-4mm}
% \end{figure}

We first present the results of image reconstruction in the Cityscape datasets, whose results are shown in Table \ref{tab:cap}. Particularly, we compare the perceptual metrics, including LPIPS and Fréchet inception distance (FID) \cite{soloveitchik2021conditional} for different methods. 
From the results,
our Diff-GO$^\text{n}$ achieves a superior LPIPS score of 0.1953 and an FID score of 54.66, outperforming prior methods such as Diff-GO+ and GESCO.
Such improvement is a direct consequence of
the restricted noised latent representation made
possible by the NB. In addition, compared with several image generation frameworks, such as SPADE, OASIS, and SMIS, our method also delivers significantly better perception
quality in terms of lower 
LPIPS and FID scores, further validating its effectiveness
in image reconstruction. Additionally, as shown in Table \ref{tab:comparison} shows strong downstream task performance for Diff-GO$^\text{n}$. These results also 
illustrate the limited ability of traditional generative AI to balance high-fidelity reconstructions with computational efficiency.

\begin{figure}[t]
%\vspace*{-5mm}
\centerline{\includegraphics[width=0.28\textwidth]{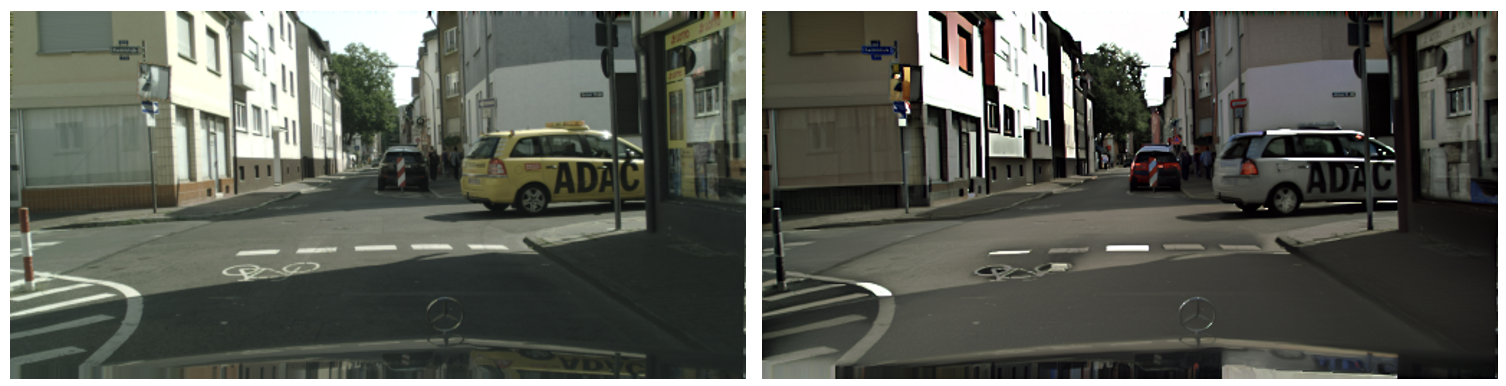}}
\vspace*{-3mm}
\caption{(Left) Image intended for transmission (TX); (Right) Reconstructed image at receiver (RX).}

\label{fig:reconstructed_images}
\vspace*{-3mm}
\end{figure}

\begin{table}[t]
  \begin{center}
    \caption{ Comparison of LPIPS and FID scores for different generative methods. The Diff-GO$^\text{n}$ (Our Method) demonstrates superior performance with the lowest LPIPS and FID scores, indicating higher perceptual similarity and realism in generated images.} \label{tab:cap}
    \begin{tabular}{|c|c|c|}
      \hline
      \textbf{Method} & \textbf{LPIPS} $\downarrow$ & \textbf{FID} $\downarrow$ \\
      \hline
      SPADE\cite{SPADE}  & 0.546 & 103.24 \\
      CC-FPSE\cite{CC-FPSE} & 0.546 & 245.9 \\
      SMIS \cite{SMIS}& 0.546 & 87.58 \\
      OASIS\cite{OASIS} & 0.561 & 104.03 \\
      SDM\cite{SDM} & 0.549 & 98.99 \\
      GESCO\cite{GESCO} & 0.591 & 83.74 \\
      Diff-GO(n=20) (250K) & 0.3206 & 74.09 \\
      Diff-GO(n=50) (250K) & 0.2697 & 72.95 \\
      Diff-GO(n=100) (250K) & 0.2450 & 68.59 \\
      Diff-Go+ (W = 64, L = 32) (250K) & 0.2231 & 60.93 \\
      Diff-Go+ (W = 128, L = 32) (250K) & 0.2126 & 58.20 \\
      \textbf{Diff-GO$^\text{n}$ (Our Method)(100K)} & \textbf{0.1953} & \textbf{54.66} \\
      % \textbf{NB-Diff-GO (Our Method)} & \textbf{0.2191} & \textbf{55.85} \\
      \hline
    \end{tabular}
  \end{center}
\end{table}
% \vspace*{-1mm}

\begin{table}[t]
  \begin{center}
    \caption{Object Detection (mIoU) and Depth Estimation (RMSE) Comparison on NB and Other Methods.} 
    \label{tab:comparison}
    \begin{tabular}{|c|c|c|c|}
      \hline
      \textbf{Method} & \textbf{Person} $\uparrow$ & \textbf{Car} $\uparrow$  & \textbf{Depth} $\downarrow$ \\ 
      \hline
       \textbf{Diff-GO$^\text{n}$} & \textbf{0.6919} & \textbf{0.7459}  &  \textbf{5.7552} \\
      Diff-GO+ & 0.6681 & 0.7457 & 5.7903 \\
      Diff-GO & 0.6241 & 0.7459 & 5.9884 \\
      GESCO & 0.5551 & 0.6644  & 7.2483 \\
      \hline
    \end{tabular}
  \end{center}
  \vspace*{-5mm}
\end{table}

  \vspace*{-1.5mm}
\subsection{Convergence Speed}
To illustrate the benefits of NB-restricted
noise sampling, we compare the convergence speed with traditional DDPM in Fig. \ref{fig:convergence} by
varying the pseudo-random NB seed values. 
With a lagging baseline, Diff-GO$^\text{n}$ shows faster convergence and superior reconstruction quality at earlier training terminations. By checkpoint 100k, every 
NB formation of Diff-GO$^\text{n}$ stabilizes at a substantially lower LPIPS score than the baseline,
clearly highlighting the efficiency of our NR-FD design in 
achieving efficient training and better perceptual quality.

\vspace*{-1.2mm}

\subsection{Effect of NR-FD and Noise Bank}
We now provide an ablation study on other benefits of NR-FD design through noise bank pre-sampling.

\subsubsection{Bandwidth Cost}
In traditional DDPM, the noised latent representation sent to the receiver requires a 384KB payload. Diff-GO+ reduces this cost using a Vector Quantized Variational Autoencoder (VQ-VAE), yielding a 4KB uncompressed payload for high-quality images \cite{Diff-go+}. In contrast, the new Diff-GO$^\text{n}$ method significantly minimizes the noised latent payload by encoding it with a single index. For instance, with an NB size of 1000, only 10 bits are needed for noise transmission.

% Here, 
% the new Diff-GO$^\text{n}$ method continues to drop
% the payload by encoding only the sampled
% noise index. For an NB size of 1000, only 10 bits
% are needed. 

\begin{figure}[t]
\centerline{\includegraphics[width=0.33\textwidth]{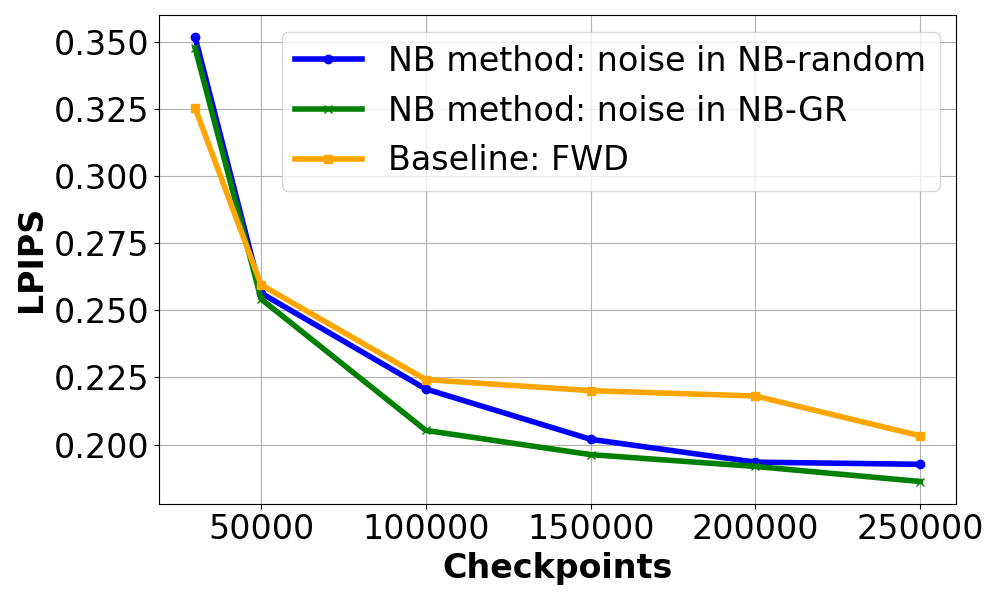}}
\vspace*{-5mm}
\caption{Comparison of Different Noise Initializations: a) NB-random (randomly selected noise); b) NB-GR (optimized noise bank); c) FWD (DDPM). }
\vspace*{-0.5mm}
\label{fig:Denoising forwad}
\end{figure}

\subsubsection{Robustness of NR-FD}
We now show the benefits resulting from the restricted noise. As illustrated in
Fig.~\ref{fig:Denoising forwad}, the basic 
NB method (NB-GR) clearly performs well. 
Interestingly, even the NB-random
method, which randomly selects a noise index to simulate
data corruption during transmission can
outperform the traditional DDPM. 
This test result demonstrates the robustness of 
the proposed Diff-GO$^\text{n}$ system
against weakly protected noised latent representation
under non-ideal channels and bit errors.

\subsubsection{Impact of NB Size}
\begin{figure}[t]
%\vspace*{-4mm}
\centerline{\includegraphics[width=0.28\textwidth]{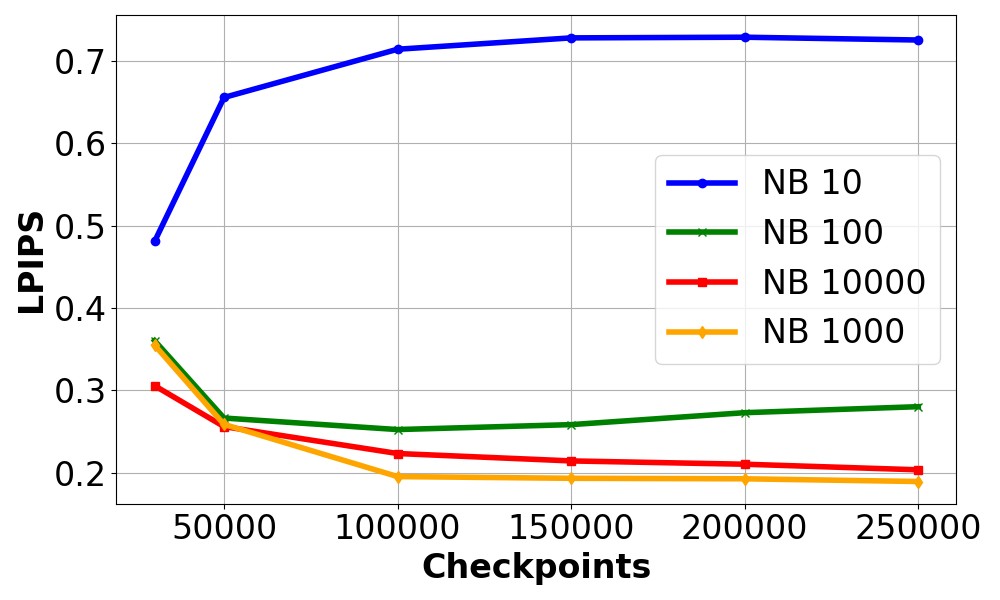}}
\vspace*{-4mm}
\caption{Effect of the Size of the Noise Bank}
\label{fig:NB size}
\vspace*{1mm}
\end{figure}

In our framework, the NB size plays a crucial role in reconstruction quality. As shown in Figure \ref{fig:NB size}, smaller Noise Bank sizes, such as size 10 (``NB-10''), 
lead to poorer performance since the model struggles to learn the structured noise effectively. Conversely, when the Noise Bank is very large, such as 10000 (``NB-10000''), the model starts to behave like the baseline DDPM with unlimited noise sampling space, losing the advantages of structured noise learning. Empirically, we found that an NB size of 1000 provides the best results for image reconstruction by
striking a good balance. In practical scenarios, 
we may optimize the NB size by utilizing a validation dataset and offer scalability to different applications.

\subsubsection{Effect of NR-FD}
\begin{figure}[t]
\vspace*{-1mm}
\centerline{\includegraphics[width=0.45\textwidth]{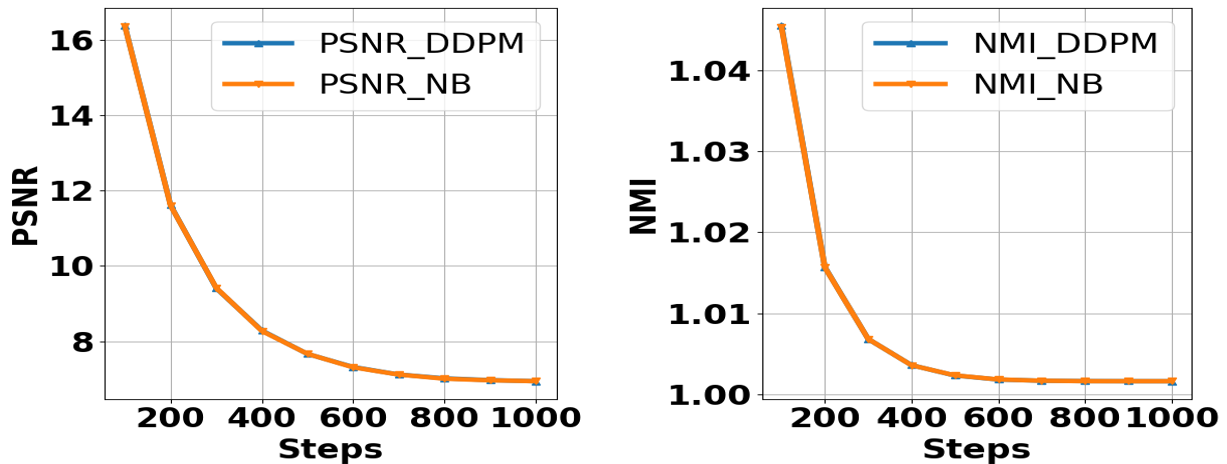}}
\caption{(Left) Normalized Mutual Information: DDPM-FD vs NR-FD; (Right) PSNR: DDPM-FD vs NR-FD}
\label{fig:fd}
\vspace*{1mm}
\end{figure}

One important consideration in Diff-GO$^n$ is whether NR-FD preserves the same low amount of original image information after forward diffusion as the baseline DDPM forward diffusion method. In the experimental results 
shown in Fig. \ref{fig:fd}, we compare the PSNR of
the resulting data from forward diffusion. 
The comparison shows that NR-FD behaves similarly to DDPM. Both methods degrade image quality at a similar rate as noise is progressively added in each step.
The comparison indicates that both NB-FD and DDPM
retain almost similar amounts of original information in each
forward diffusion step despite the use of a pre-sampled NB for
noise restriction. This indicates that NR-FD effectively tracks the traditional DDPM throughout its diffusion process.

\vspace*{-1.5mm}
\section{Conclusion}
\vspace*{-1mm}
This work develops a novel diffusion
framework Diff-GO$^\text{n}$ for goal-oriented 
communications (GO-COM). Our pre-sampled NB supports the new Noise-Restricted Forward Diffusion 
(NB-FD) concept. We further provide an early stopping criterion to overcome the high computational cost in traditional diffusion-based GO-COM designs. 
The structured noise patterns significantly reduce the payload overhead and bandwidth consumption while still achieving satisfactory data regeneration at 
the receiver output.
Our experimental results demonstrate the power of 
Diff-GO$^\text{n}$ in terms of faster convergence, better reconstruction quality, and robustness under various 
conditions. For future studies, we plan to investigate the efficient derivation of noise banks and lightweight
simplification of the underlying deep-learning architecture. 

\vspace{-1.5mm}
\bibliographystyle{IEEEtran}
\bibliography{main}

\begin{thebibliography}{10}
\providecommand{\url}[1]{#1}
\csname url@samestyle\endcsname
\providecommand{\newblock}{\relax}
\providecommand{\bibinfo}[2]{#2}
\providecommand{\BIBentrySTDinterwordspacing}{\spaceskip=0pt\relax}
\providecommand{\BIBentryALTinterwordstretchfactor}{4}
\providecommand{\BIBentryALTinterwordspacing}{\spaceskip=\fontdimen2\font plus
\BIBentryALTinterwordstretchfactor\fontdimen3\font minus \fontdimen4\font\relax}
\providecommand{\BIBforeignlanguage}[2]{{%
\expandafter\ifx\csname l@#1\endcsname\relax
\typeout{** WARNING: IEEEtran.bst: No hyphenation pattern has been}%
\typeout{** loaded for the language `#1'. Using the pattern for}%
\typeout{** the default language instead.}%
\else
\language=\csname l@#1\endcsname
\fi
#2}}
\providecommand{\BIBdecl}{\relax}
\BIBdecl

\bibitem{9349624}
W.~Jiang, B.~Han, M.~A. Habibi, and H.~D. Schotten, ``The road towards 6g: A comprehensive survey,'' \emph{IEEE Open Journal of the Communications Society}, vol.~2, pp. 334--366, 2021.

\bibitem{10570351}
A.~Mostaani, T.~X. Vu, H.~Habibi, S.~Chatzinotas, and B.~Ottersten, ``Task-oriented communication design at scale,'' \emph{IEEE Transactions on Communications}, pp. 1--1, 2024.

\bibitem{Diff-go}
A.~Wijesinghe, S.~Zhang, S.~Wanninayaka, W.~Wang, and Z.~Ding, ``Diff-go: Diffusion goal-oriented communications to achieve ultra-high spectrum efficiency,'' \emph{arXiv preprint arXiv:2312.02984}, 2023.

\bibitem{o2017introduction}
T.~O’shea and J.~Hoydis, ``An introduction to deep learning for the physical layer,'' \emph{IEEE Transactions on Cognitive Communications and Networking}, vol.~3, no.~4, pp. 563--575, 2017.

\bibitem{SPADE}
T.~Park, M.-Y. Liu, T.-C. Wang, and J.-Y. Zhu, ``Semantic image synthesis with spatially-adaptive normalization,'' \emph{Proc. IEEE Conf. Computer Vision and Pattern Recog. (CVPR)}, 2019.

\bibitem{OASIS}
E.~Schonfeld, B.~Schiele, and A.~Khoreva, ``Object-aware semantic image synthesis,'' \emph{arXiv preprint arXiv:2011.12026}, 2020.

\bibitem{CC-FPSE}
Q.~Chen and V.~Koltun, ``Contextual conditioning for generative adversarial networks,'' \emph{arXiv preprint arXiv:1706.07122}, 2017.

\bibitem{croitoru2023diffusion}
F.-A. Croitoru, V.~Hondru, R.~T. Ionescu, and M.~Shah, ``Diffusion models in vision: A survey,'' \emph{IEEE Transactions on Pattern Analysis and Machine Intelligence}, vol.~45, no.~9, pp. 10\,850--10\,869, 2023.

\bibitem{GESCO}
E.~Grassucci, S.~Barbarossa, and D.~Comminiello, ``Generative semantic communication: Diffusion models beyond bit recovery,'' \emph{arXiv preprint arXiv:2306.04321}, 2023.

\bibitem{Diff-go+}
A.~Wijesinghe, S.~Zhang, S.~Wanninayaka, W.~Wang, and Z.~Ding, ``Diff-go+: An efficient diffusion goal-oriented communication system with local feedback,'' \emph{Authorea Preprints}, 2024.

\bibitem{DDPM}
J.~Ho, A.~Jain, and P.~Abbeel, ``Denoising diffusion probabilistic models,'' \emph{Adv. Neural Info. Proc. Systems}, vol.~33, pp. 6840--6851, 2020.

\bibitem{improved}
A.~Q. Nichol and P.~Dhariwal, ``Improved denoising diffusion probabilistic models,'' in \emph{International conference on machine learning}.\hskip 1em plus 0.5em minus 0.4em\relax PMLR, 2021, pp. 8162--8171.

\bibitem{zhang2018unreasonable}
R.~Zhang, P.~Isola, A.~A. Efros, E.~Shechtman, and O.~Wang, ``The unreasonable effectiveness of deep features as a perceptual metric,'' in \emph{Proc. IEEE Conf. Computer Vision and Pattern Recog. (CVPR)}, 2018, pp. 586--595.

\bibitem{radius}
Y.~Zhu, Y.~Wu, Z.~Deng, O.~Russakovsky, and Y.~Yan, ``Boundary guided learning-free semantic control with diffusion models,'' \emph{Advances in Neural Information Processing Systems}, vol.~36, 2024.

\bibitem{Cityscapes}
M.~Cordts, M.~Omran, S.~Ramos, T.~Rehfeld, M.~Enzweiler, R.~Benenson, U.~Franke, S.~Roth, and B.~Schiele, ``The cityscapes dataset for semantic urban scene understanding,'' in \emph{Proc. IEEE Conf. Computer Vision and Pattern Recog. (CVPR)}, 2016.

\bibitem{soloveitchik2021conditional}
M.~Soloveitchik, T.~Diskin, E.~Morin, and A.~Wiesel, ``Conditional frechet inception distance,'' \emph{arXiv preprint arXiv:2103.11521}, 2021.

\bibitem{SMIS}
T.~Park, M.-Y. Liu, T.-C. Wang, and J.-Y. Zhu, ``Semantic image synthesis with spatially-adaptive normalization,'' \emph{Proc. IEEE Conf. Computer Vision and Pattern Recog. (CVPR)}, 2019.

\bibitem{SDM}
J.~Ho, A.~Jain, and P.~Abbeel, ``Denoising diffusion probabilistic models,'' \emph{arXiv preprint arXiv:2006.11239}, 2020.

\end{thebibliography}

% \section*{\textcolor{red}{We should put supplemental materials with extensive tests and intermediate results}}
\end{document}